%% file: corona.tex
\documentclass[epsf]{aastex}
\usepackage{apjfonts}
\usepackage[onecolumn]{emulateapj5}
\usepackage{graphicx} 
\usepackage{color} 
\singlespace
 
\newcommand{\be}{  \begin{eqnarray} }
\newcommand{\ee}{  \end{eqnarray} }

\def\spose#1{\hbox to 0pt{#1\hss}}
\def\lta{\mathrel{\spose{\lower 3pt\hbox{$\mathchar"218$}}
     \raise 2.0pt\hbox{$\mathchar"13C$}}}
\def\gta{\mathrel{\spose{\lower 3pt\hbox{$\mathchar"218$}}
     \raise 2.0pt\hbox{$\mathchar"13E$}}}

\begin{document}

\shorttitle{Coronal-Powered SNe and GRBs}
\title{Supernovae and gamma-ray bursts powered by hot 
neutrino-cooled coronae}
\author{ Enrico Ramirez-Ruiz\altaffilmark{1,2} and 
Aristotle Socrates\altaffilmark{3,4}}
\altaffiltext{1}{Institute for Advanced Study, Einstein Drive,
Princeton, NJ 08540: enrico@ias.edu}
\altaffiltext{2}{Chandra Fellow}
\altaffiltext{3}{Department of Astrophysical Sciences, Princeton 
University, Peyton Hall-Ivy Lane, Princeton, NJ 08544: 
socrates@astro.princeton.edu}
\altaffiltext{4}{Hubble Fellow}
\begin{abstract}

Cosmological explosions such as core-collapse supernovae (SNe) and
gamma-ray bursts (GRBs) are thought to be powered by the rapid
conversion of roughly a solar mass' worth of gravitational binding
energy into a comparatively small amount of outgoing observable
kinetic energy.  A fractional absorption of the emitted neutrinos, the
particles which carry away the binding energy, by the expelled matter
is a widely discussed mechanism for powering such explosions.
Previous work addressing neutrino emission from core-collapse like
environments assumes that the outgoing neutrino spectrum closely
resembles a black body whose effective temperature is determined by
both the rate of energy release and the surface area of the entire
body. Unfortunately, this assumption minimizes the net efficiency for
both neutrino-driven explosion mechanisms.  Motivated by this fact, we
qualitatively outline a scenario where a hot corona deforms the
neutrino spectrum away from that of a cool thermal emitter.  Our
primary result is that in principle, a coronal-driven explosion
mechanism can enhance the net efficiency of neutrino-driven SNe and
GRBs by more than an order of magnitude.
\end{abstract}


\section{the basic idea and initial estimates}

Core-collapse supernovae (SNe) are thought to be powered by thermal
neutrinospheric emission coupling, however weakly, to an outgoing
prompt shock (Bethe 1990, Herant et al. 1994, Janka 2001).  After
decades of theoretical and detailed numerical analysis, the robustness
of the prompt mechanism is still in doubt (Lieberndorfer 2001 and
references therein).  It seems as though some additional physical
process is required.

A popular model for powering gamma-ray bursts (GRBs) involves
hyper-Eddington accretion, where the release of energy is mediated by
neutrino emission, onto a stellar mass black hole (Eichler et
al. 1989; Woosley 1993).  In the evacuated region above the black
hole, neutrinos and their anti-particles annihilate into
electron-positron pairs and a relativistic fireball is formed (Woosley
1993; Goodman et al. 1987).  Preliminary studies of this mechanism
indicate that the efficiency of converting neutrino radiation into
pairs is somewhat smaller than that needed to power a long-duration
($t> 2\;$s) burst (DiMatteo et al. 2002, Rosswog \& Ramirez-Ruiz 2002).

The most common approach in circumventing the shortcomings of
neutrino-powered explosions in both cases is to increase the overall
neutrino luminosity.  In the SNe case, the energy budget $\sim$ a
few$\times 10^{53}$erg is capped by the mass of the star for a
standard equation of state.  Therefore, the proto-neutron star's
(PNSs) luminosity can only be increased by reducing the
Kelvin-Helmholtz time which is set, to first approximation, by
neutrino diffusion.  Convection in PNSs beneath the neutrinosphere
might play a prominent role in increasing the star's luminosity at
early times, but it does so by only a modest factor (Burrows 1987).
In terms of GRB accretion disks, the net thermal luminosity can be
increased by increasing the accretion rate.  Upon doing so, the
neutrino trapping radius increases and a limiting luminosity is
reached, leading to low annihilation efficiencies (DiMatteo et
al. 2002).

In this work, we take a different approach.  Instead of finding ways
to increase the total output of gravitational binding energy during
the explosion epoch, we consider deformations of the neutrino spectrum
away from that of a single optically thick thermal emitter.  For SNe,
the most important energy deposition process above the neutrinosphere
is mediated by the capture of electron-type neutrinos onto free
nucleons i.e., 
\be \nu_e+n &\rightarrow & e+p\\ {\bar\nu_e}+p &
\rightarrow & e^++n.  
\ee 
For both of these reactions, the deposition
rate can be written as 
\be Q^+_{\nu,N}=\sigma_0\,Y_N n\,
\frac{L_{\nu}}{A}\left<E^2_{\nu}\right>
\label{dep1}
\ee where $\sigma_0=4.5\times 10^{-44}{\rm cm^2}\,{\rm MeV^{-2}}$,
$n$, $Y_N$, $L_{\nu}$, $A$, and $\left<E^2_{\nu}\right>$ is a
characteristic weak interaction cross section, number density of free
baryons, neutron (proton) fraction, electron (anti-electron) neutrino
luminosity, surface area of the absorbing region, and
spectrum-averaged square of the electron (anti-electron) neutrino
energy, respectively.  For GRBs powered by hyper-Eddington accretion
flows the relevant deposition process is neutrino annihilation \be
\nu_e+{\bar\nu_e} \rightarrow e^- + e^+ \ee with a corresponding
deposition rate given by \be
Q^+_{\nu,{\bar\nu}}={\bar\sigma_0}\frac{L^2_{\nu}}{A^2} \left<
E_{\nu}\right>\,\zeta
\label{dep2}
\ee where ${\bar\sigma_0}=3K\,G^2_F/4$ is another characteristic weak
interaction cross section per unit energy squared, $G^2_F=5.29\times
10^{-44}\,{\rm cm^2}\,{\rm MeV^{-2}}$ is the Fermi constant, $K$ is a
phenomenological electro-weak parameter usually taken to be $0.1-
0.2$, and $A$ is the surface area of the absorbing region.  In the
above expression, we assumed, for the sake of compactness, that the
luminosity spectrum of both neutrino and anti-neutrinos are identical,
and therefore $\left<E_{\nu} \right>$ is simply the mean neutrino
energy weighted over the neutrino spectrum.\footnote{Implicitly, we
are approximating that $\left<E_{\nu}
\right>^2=\left<E^2_{\nu}\right>$.  Of course, this introduces errors
at less than the order unity level for a thermal emitter at fixed
luminosity.  However, the shape of the spectrum must carefully be
taken into account if an accurate computation of eqs. (\ref{dep1}) and
(\ref{dep2}) is desired when considering a coronal-driven explosion
mechanism.  Regardless, at this preliminary level, we ignore such
details for the sake of clarity.}  Also, the multiplicative factor
$\zeta$ takes into account the geometry of the emitting region.

Eqs. (\ref{dep1}) and (\ref{dep2}) state that for a fixed amount of
energy release i.e., $L_{\nu}$, the efficiency of producing an
explosion in either case is {\it minimized} when the spectrum
resembles a pure single black body since black bodies maximize the
number of emitted quanta while minimizing the mean energy per quanta for a
given energy flux.\footnote{Throughout this paper, when we refer to a
thermal single temperature black body emitter, we necessarily 
imply that the neutrino chemical potential is zero.  However, 
deviations from zero chemical potential do occur as a result of  
transfer effects near the neutrinosphere (e.g. Keil et al. 2003). 
The magnitude to these
deviations to the mean neutrino energy are relatively small compared
to those discussed here.}  It follows, that a 
significant spectral deviation
from that of a single black body in eqs. (\ref{dep1}) and (\ref{dep2}) 
enhances the explosion mechanism for core-collapse SNe
and GRBs.  From here on, we assume a hot diffuse region that lies
directly above the cool dense flow, in other words a ``corona,'' is
responsible for producing any spectral deformations.  We may simplify
our discussion by separating the neutrino energy spectrum into a soft
component $L^s_{\nu}$, which emanates from the cool dense regions, and
hot coronal component $L^c_{\nu}$.  Our simple parameterization allows
us to write
\be
L_{\nu}\left<E^n_{\nu}\right>=L^{s}_{\nu}\left<E^n_{\nu}\right>_{s}
+L^{c}_{\nu}\left<E^n_{\nu}\right>_{c}
\label{Lsplit}
\ee  
where $n=1,2$ for the neutrino annihilation and capture, respectively.
If a fraction $f$ of the energy release is mediated by the corona, then 
\be
L_{\nu}\left<E^n_{\nu}\right>=L_{\nu}\left[\left(1-f\right)\left<E^n_{\nu}
\right>_{s}+f\left<E^n_{\nu}\right>_{c}\right].  
\label{Lsplit2}
\ee 
Considering the SNe mechanism (n=2), if $f=0.1$ and if the average
energy of a non-thermal or ``coronal'' neutrino is $\sim 10$ times
that of the average thermal neutrino, then the deposition rate from
neutrino absorption given by eq. (\ref{dep1}) is $\sim 10$ times
larger than the minimum thermal case.  In order to obtain such a large
increase in energy deposition for GRBs, the neutrino spectrum must be
deformed to a much more extreme degree.  Note that the above scalings
applies to momentum deposition as well or in other words, the
radiation force.  That is, if energy deposition increases by some
factor due to non-thermal coronal emission, then the neutrino
radiation force increases accordingly by the same amount.

\section{the existence of a coronae in hot dense matter}

The dynamics and energy release mechanisms which govern cool
dense turbulent stratified astrophysical flows are not well understood.
Regardless, we now qualitatively discuss some of the basic
requirements for coronal formation in neutrino-cooled systems while
keeping in mind the enormous level of uncertainty attached to
our assumptions.   

Two physical quantities determine whether or not coronal neutrino
emission is able to drive an explosion: the fraction of gravitational
power $f$ released in the corona and the temperature $T$ at which the
neutrinos are emitted.  The first quantity $f$ is determined by the
complicated physical mechanisms that determine the structure and
emitted radiation spectra of these intense turbulent flows.
At the same time, the coronal temperature $T$ is calculated by
specifying the amount of mass and volume in which the fraction $f$ of
the energy release is deposited.

\subsection{Energetics}
  
If all of the radiative energy release occurs in the relatively opaque
dense regions, then $f=0$ and conversely, if all of the energy release
is mediated by a hot optically thin region, then $f=1$.  That is, 
$f$, is the {\it mechanical} energy that is deposited in
the diffuse upper atmosphere of a PNS or hyper-Eddington accretion flow.
For PNSs in particular, the fraction $f$ can be even more 
precisely defined as the ratio that 
quantifies the amount of mechanical energy that passes through
the neutrinosphere of the opaque nascent star.

An obvious candidate for generating interesting values of $f$ are the
magnetic field structures generated by the putative turbulence in
these flows.  Below, we assess the viability of turbulent magnetic
energy transport for both PNSs and hyper-Eddington accretion flows.
Figure 1 illustrates the general features of a coronal-driven 
explosion mechanism for both SNe and GRBs.

\begin{figure}[t!]
\begin{center}
\input{layer.pstex_t}
\caption{Sketch of a neutrino-driven explosion. 
For SNe, the outflow takes the form of a massive sub-relativistic shock, 
while for GRBs the outflow is a relativistic high entropy wind.  
Binding energy, the given system's ultimate energy source, is stored in
the cool, dense, turbulent layer -- where nearly all of the mass 
resides.  The turbulence serves as the source of magnetic energy and 
is generated passively in a PNS and actively is an accretion disk, perhaps
by the MRI (Balbus \& Hawley 1991).  When magnetic structures
rise to and above the surface of the cool layer, they release their 
energy in relatively small amounts of mass and volume.  If the magnetic
energy is deposited uniformly at a given radius, then the corona
is endowed with a scale height $H$ as discussed in 
\S\ref{ss:opacity}.  Alternatively, the magnetic dissipation may occur 
inhomogenously, much like the Sun.  In this case, the relevant scale for 
magnetic energy deposition is characteristic size of a hot spot, $R_s$ 
which is discussed in  \S\ref{ss:opacity} as well.            }
  
\end{center}
\label{f:model}
\end{figure}
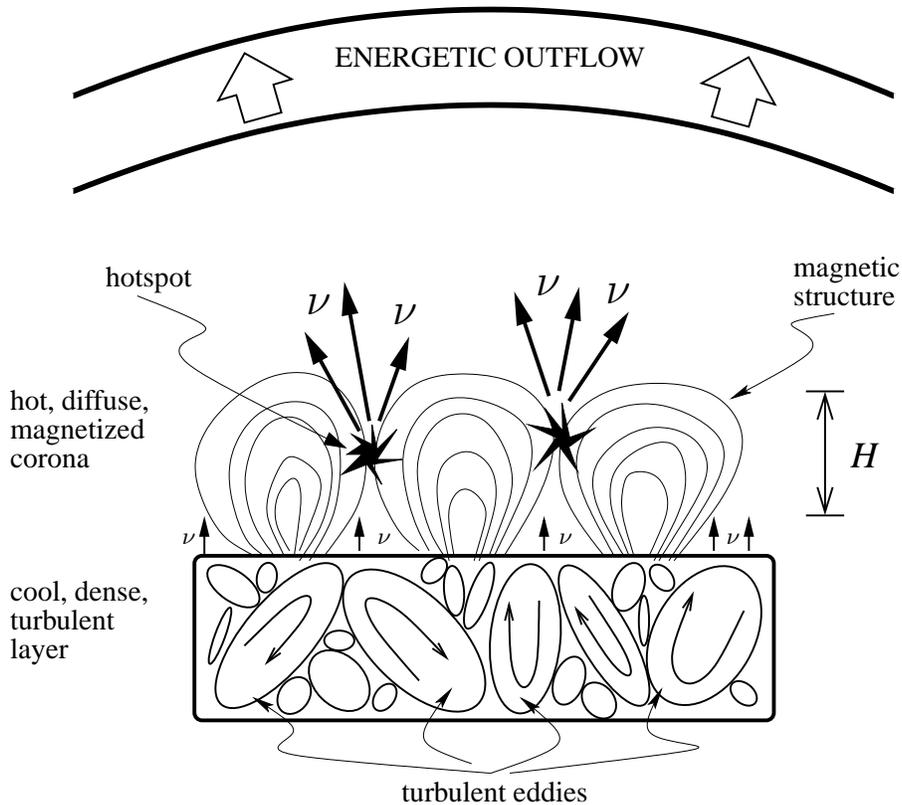
\subsubsection{Convective Proto-Neutron Stars}

PNSs are thought to convective for a variety of reasons immediately
after their birth (Thompson \& Murray 2001, hereafter TM01; Keil,
Janka, \& Mueller 1996).  When compared to their photon-limited main
sequence progenitors, the convective zones of PNSs are $\sim 10^4$
times more energetic with respect to their own binding energy,
implying that they are ideal sites for dynamo activity (Duncan \&
Thompson 1992; Thompson \& Duncan 1993, hereafter TD93).  At this
preliminary level, we assume that the PNS in question is not rapidly
rotating such that magnetic field generation occurs in the absence of
a net source of helicity, similar to the field generation process
found in the Sun's inactive regions.  However, unlike the solar
convection zone, the turbulent diffusivity due to convection decreases
outwards.  Therefore, the combined actions of buoyancy and turbulent
transport conspire to ``pump'' turbulent magnetic field structures
towards the stellar surface (TM01; Socrates et al. 2004).  We may
approximate the strength of the rms turbulent magnetic field by
writing \be {B^2_{\rm turb}}/{4\pi}\sim\epsilon_B\rho v^2_c \ee where
$v_c$ is the convective velocity and $\epsilon_B$ is an efficiency
factor.\footnote{Analysis of solar $p-$mode frequency shifts reveal
that convective motions beneath the photosphere maintain significant
turbulent magnetic stresses such that $\epsilon_B\sim 0.1$ down to
several pressure scale heights beneath the photosphere (Goldreich et
al. 1991).}  By estimating the value of the convective energy flux
transmitted to the neutrinosphere via mixing length theory \be
F_c\sim\rho v^3_c, \ee implies that the fraction $f\simeq\epsilon_B$
since turbulent magnetic structures rise and are mixed upward at a
velocity $\sim v_c$.  Thus, for reasonable choices of magnetic field
strength such that $\epsilon_B\sim 0.1$, a large amount of the stars
binding energy, $\sim$ a few $\times 10^{52}\,{\rm ergs}$, may be
dissipated in a highly magnetized optically thin corona.  This can
only be the case if the star is convective for a significant fraction
of its Kelvin-Helmholtz time.

TD93 noticed that if only a small fraction of the turbulence generated
magnetic energy is deposited in an outgoing shock, then a supernova
can be powered by magnetic energy alone.  If the dynamics and energy
release mechanisms of PNS convection zones are similar to those
observed in the Sun, for example, then a magnetically powered
explosion is highly unlikely since $f$ would be too small (TD93).  
However, there are significant physical differences
between the convection zones of main sequence stars and those of PNSs,
some of which have already been mentioned.  Another important
difference is that the ratio ${\mathcal R}$ of kinetic energy of the
turbulence to the binding energy of the turbulent layer itself is
quite large in a PNS relative to that of a main sequence star.  For
PNSs, ${\mathcal R}\sim 10^{-3}$, whereas this ratio is closer to
$\sim 10^{-5}-10^{-6}$ in the Sun.  Interestingly, if relativistic
accretion flows resemble a violently turbulent layer, then for certain
models of accretion, ${\mathcal R}$ may approach values of $10^{-2}$
or larger for X-ray binaries and active galactic nuclei (see also Thompson
1994).
                                                                            
\subsubsection{Neutrino-Powered Accretion Disks}

Our knowledge regarding the structure and energy release mechanisms of
hyper-Eddington accretion flows, which may power GRBs, does not enjoy
the same relative level of certainty as their core-collapse SNe
counterparts.  However, the likely physical parameters of both systems
e.g., density, temperature, surface gravity, and surface area of the
luminous neutrino emitting region are quite similar.  For our work,
another important similarity is that both systems are highly turbulent
and magnetized.

Since a predictive theory of multi-phased relativistic accretion is
currently not available, we resort to observations and analysis of
photon-powered black hole sources for guidance.  In the case of Cyg
X-1 and other black hole X-ray binaries (BHXRBs), the radiation
spectrum in the low/hard state is dominated by a hot coronal
component, which is most likely generated by thermal comptonization.
From this fact alone, we argue phenomenologically that the radiation
spectrum emanating from hyper-Eddington accretion flows may possibly
resemble that of Cyg X-1 and other BHXRBs in the low/hard state, 
implying a value for $f\sim 1$ during some fraction of the disk's
viscous time. 

The above assertion is not completely devoid from all theoretical
reasoning. For example, one model of the X-ray spectrum of BHXRBs and
Seyfert 1 Galaxies invokes the transport of magnetic energy away from
a cool dense disk into a hot diffuse corona such that $f\sim 1$.  This
notion reproduces the observed spectral features while modeling the
corona as a sum of compact magnetic flares, perhaps generated by the
MRI, which produce mildly relativistic pair-dominated outflows
(Beloborodov 1999, see also Galeev, Rosner \& Vaiana 1979 and Miller
\& Stone 2000).  Therefore, the qualitative {\it physical} argument
that further motivates the possibility that hyper-Eddington accretion
flows take on large values of $f$ is that they are magnetically
turbulent in a manner similar to that of sub-Eddington accretion
flows.  It follows, that if a corona is responsible for the majority
of the radiative energy release, then $\sim$ a few $\times 10^{53}{\rm
ergs}$ of gravitational binding energy per $M_{\odot}$ accreted is
emitted in a hot diffuse magnetized phase.

\subsection{Constraints from Neutrino Absorption Opacity}\label{ss:opacity}

The neutrino continuum scattering and absorption opacity increases
with energy.  This microphysical difference between neutrino-cooled
and photon-cooled systems does not lead to qualitative differences
so long as the flow is optically thick.  However, the structure 
of diffuse optically thin
regions heavily depends on how efficiently matter can cool and
therefore, an increasing absorptivity (and emissivity) with particle
energy places different constraints on coronal structure compared to
the photon-cooled case.

For PNS envelopes and hyper-Eddington accretion flows, pair capture
and annihilation serve as the dominant sources of opacity.  The
emissivity for later process may be written in terms of a local
temperature 
\be Q^-_{\nu,{\bar\nu}}\simeq 5\times
10^{33}T^9_{11}\,{\rm ergs}\, {\rm cm^{-3}\,s^{-1}}
\label{Qann}
\ee
where $T_{11}=T/10^{11}\;$K and the analogous expression for the pair capture
process reads
\be
Q^-_{eN}\simeq 9\times 10^{33}\rho_{10}T^6_{11}\,{\rm ergs}\,
{\rm cm^{-3}\,s^{-1}}
\label{Qcap}
\ee and similarly, the corresponding absorption optical depths for a
characteristic length scale $H$ (in cm) are respectively \be
\tau^a_{\nu,{\bar\nu}}\simeq
\frac{Q^-_{\nu,{\bar\nu}}\,H}{4\left(7/8\right)\sigma T^4}\simeq
2.5\times 10^{-7}\,T^5_{11}\,H
\label{tau1}
\ee
and
\be
\tau^a_{eN}\simeq\frac{Q^-_{\nu,{eN}}\,H}{4\left(7/8\right)\sigma
T^4}\simeq 4.5\times 10^{-7}\,\rho_{10}\,T^2_{11}\,H
\label{tau2}
\ee 
(di Matteo et al. 2002).\footnote{The temperature and density
dependencies of Eqs. (\ref{Qann}) and (\ref{Qcap}) can be understood
as follows.  The absorption opacity in both cases are $\propto T^2$.
However, the number density of the absorbing particle species in the
pair annihilation case is $\propto T^3$ while in the pair capture case
the number density of free baryons is $\propto\rho$.  By recalling
that the energy density of relativistic electrons and positrons are
$\propto T^4$, the temperature and density dependencies of both
processes are recovered.}  Note that we have assumed that all nuclei
are completely dissociated into free baryons.

In order to determine the structure of a neutrino-cooled corona, the
(mechanical and radiative) energy deposition rate per unit mass must
be specified -- a task that is well outside the scope of this work.
In order to proceed, we make simplifying assumptions with respect to
the overall geometry of the corona and mechanical energy deposition
profile.  The simplest configuration requires a constant density,
temperature, scale height, and mechanical energy deposition per unit
mass i.e., a corona resembling a uniform plane-parallel slab.  From
eqs. (\ref{dep1}) and (\ref{dep2}), we see that the explosion
efficiency is highly dependent on the energy, and thus the temperature
at which neutrinos are emitted. Our corona must be both optically thin
\be \tau^a_i<1
\label{condition1}
\ee
and satisfy radiative equilibrium
\be
f\,L_{\nu}=Q^-_i\,V
\label{condition2}
\ee where $i$ represents either pair annihilation or capture and $V$
is the volume of the slab.  For a PNS, $V=4\pi R^2_{\rm PNS}H$ while
for an accretion disk $V=2\pi R^2_dH$ where $R_{\rm PNS}$ and $R_d$
are the radius of the PNS and the characteristic disk radius,
respectively.  Condition (\ref{condition1}) enforces that the corona,
with a surface area roughly equal to that of the cool dense emitting
fluid, does not radiate at the effective black body temperature.  By
{\it fiat} the corona is a low density region and therefore, pair
annihilation is most likely the dominant source of opacity as long as
the mechanical energy is injected sufficiently far away from the cool
dense regions.  In this case, the above conditions yield a coronal
temperature \be T\simeq
10^{11}\frac{f^{1/4}_{-1}L^{1/4}_{53}}{\tau^{1/4}_{-2}R^{1/2}_{6.5}}\,
{\rm K}, \ee which corresponds to a coronal scale height of $H\simeq
3\times 10^4\,{\rm cm}$.\footnote{For reference, the scale height $H$
at the neutrinosphere for a PNS is of order $\sim 10^{5}$ cm (Janka
2001).} The above temperature implies an electron neutrino energy
$\sim 30-40$ MeV, which is $\sim$ a few times larger then the energy
of the electron neutrinos emitted from a newly born optically thick
PNS.

The other coronal geometry which we explore corresponds to a region
above the cool dense flow consisting of $N$ neutrino-emitting hot spots
of characteristic spot size $R_s$.  In this case, a hot spot can be
optically thin or thick.  At this preliminary level, we further assume
that all of the hot spots are either optically thin or optically thick.
In the optically thick case, radiative equilibrium over the entire
corona requires \be f\,L_{\nu}\simeq\sigma\,T^4\,N\,R^2_s, \ee where
$\sigma$ is the Stefan-Boltzmann constant, leading to a characteristic
spot temperature of \be T\simeq
10^{12}\frac{f^{1/4}_{-1}L^{1/4}_{53}}{N^{1/2}_2\,
R^{1/2}_{s,3}}\,{\rm K}.  \ee This particular coronal configuration of
optically thick hot spots yields an optical depth per spot $\tau\sim
30$, implying a radiative diffusion times $\sim \tau^2\,R_s/c\sim$ a
few $\times 10^{-5}\,{\rm s}$.  Therefore, neutrino cooling within the 
hot spot is instantaneous compared to the eddy turnover time 
$\sim 1\,{\rm ms}$ i.e., the likely timescale of magnetic 
energy injection.

For the optically thin case, condition (\ref{condition1})
must be enforced for each hot spot, while the volume $V$ of the
emitting region is now given by $V\simeq N\,R^3_s$ when calculating
radiative equilibrium.  Here, the coronal temperature is roughly 
\be
T\simeq 9\times
10^{11}\frac{f^{1/9}_{-1}L^{1/9}_{53}}{R^{1/3}_{s,1.5}N^{1/9}_2} {\rm
K}, \ee 
which marginally obeys (\ref{condition1}) for each hot spot.  The
neutrino energies quoted above are $\sim$ 3-30 times larger than that
of the thermal electron-type neutrinos emitted from the optically
thick surfaces of PNS and hyper-Eddington accretion flows.  As long as
mechanical energy deposition above the neutrinosphere is concentrated,
either homogeneously or in-homogeneously, in a small volume at low
densities, the spectral component emanating from the coronal may
significantly contribute to or even entirely dominate the explosion
mechanism.

\subsubsection{A Note on $e-\nu_e$ Comptonization}

The $e^{+/-}$ pairs in the putative corona are highly relativistic and 
have thermal energies in excess of $\left<E_{\nu}\right>_s$.  It 
follows, that the hot coronal $e^{+/-}$ pairs can cool by up-scattering the 
soft neutrinos as a result of the process
\be
e+\nu_e\rightarrow e+\nu_e,
\ee 
an effect that is particularly relevant for the 
slab geometry due to its large covering fraction.  Comptonization 
produces a non-thermal deformation to the soft spectral component.  
Incidentally, this effect is believed to deform the spectrum of 
primordial neutrinos after their 
respective decoupling from matter (Dolgov \& Fukugita 1992).  For systems  
possessing photon-cooled coronae such as Seyfert Galaxies, X-ray binaries,
and active stars, Compton up-scattering of soft seed photons
by hot electrons is often the dominant source of energy 
release.  As previously discussed, this is not necessarily the case
for neutrino-cooled coronae since the release of energy due to the
continuum emissivity becomes increasingly important for large
temperatures.  

Like photon-cooled systems (see Sunyaev \& Titarchurk 1980), the 
relative importance of $e-\nu_e$ Comptonization as a radiative energy
release mechanism is most easily quantified through a $y$--parameter
\be
y_{e\nu}\sim \frac{\Delta E_{\nu}}{E_{\nu}}\,\tau_C
\ee
where $\Delta E_{\nu}$ is the energy gain per scattering event, $E_\nu=\left<
E_{\nu}\right>_s$ is 
the energy of a soft neutrino prior to up-scattering, and $\tau_C$ is
the Compton optical depth, which is given by $\tau_C\sim \sigma_C\, n_e\,H$.
The Compton cross section $\sigma_C\sim \sigma_0\left<E_{\nu}\right>_s
\left<E_e\right>_c$, where $\left<E_e\right>_c\simeq\left<E_{\nu}\right>_c$
is the energy of a hot coronal electron.  Thus, 
$\sigma_C$ is smaller 
than the annihilation and capture cross sections by a factor $\sim 
\left<E_{\nu}\right>_s /\left<E_e\right>_c$.  The number density of Comptonizing 
leptons $n_e$
is $\propto T^3$ for high entropy radiation pressure 
dominated regions and is $\propto Y_e\, n$ for low entropy regions that are 
gas pressure dominated.  Here, $Y_e$ is the electron fraction and $n$ is the number 
density of free baryons.  A large energy shift $\Delta E_{\nu}\sim \left<E_e
\right>_c$
accompanies each scattering as long as the coronal temperature is 
significantly hotter than the cool dense region responsible for
providing the soft seed neutrinos. Interestingly, the $y$--parameter may 
be written in the following form
\be
y_{e\nu}\sim \frac{\left<E_e\right>_c}{\left<E_{\nu}\right>_s}\sigma_0\left<E_{\nu}
\right>_s\left<E_e\right>_c\,n_e\,H\sim \tau_a
\ee          
where $\tau_a$ is given by (\ref{tau1}) and (\ref{tau2}) for the radiation and 
gas pressure dominated cases or in other words, the pair annihilation and 
pair capture cases, respectively.  Even for modest values of 
$y_{e\nu}$ such that $y_{e\nu}\sim 0.1$, Compton cooling of hot electrons
plays an important role in mediating the release of a corona's energy 
budget as long as the covering fraction of the up-scattering region
is not too small.

\section{Discussion}

To our knowledge, all previous discussions regarding prompt neutrino
emission from dense core-collapse like environments assume that the
entire surface of the object in question radiates close to the black
body limit.  The effective temperature is determined by both the rate
of gravitational energy release and the surface area of the star or
disk.  Therefore, total neutrino number, for a given luminosity, is
maximized while the average energy per neutrino is minimized.  A
direct consequence of this choice is that the efficiency of a
neutrino-driven explosion mechanism for both SNe and GRBs is
minimized.  In the previous section, we put forth a qualitative
picture as to how a hot corona could alter the neutrino spectrum from
that of a pure black body.  In what follows, we briefly discuss how
significant deviations from a pure black body can alter the respective
explosion mechanisms for SNe and GRBs.

\subsection{Core-Collapse Supernovae}

In light of the overall uncertainty surrounding the thermal neutrino
powered explosion model (Bethe \& Wilson 1986) and its convective
variant (Herant et al. 1994; Burrows et al. 1995), many theorists
have resorted to alternate sources of energy, other than
gravitational, in order to power core-collapse SNe.  An obvious choice
is rotational energy, which one way or the other, is believed to lead
to magnetic field amplification, which subsequently gives rise to spin
down torques as well as an extra source of heating (cf. Thompson et al. 
2004; 2005).  However, it is
unlikely that rotation can effect the explosion mechanism unless the
initial spin period of a young neutron star is significantly in excess
of those measured in young radio pulsars.  On the other hand, vigorous
convection is likely to persist throughout the explosion epoch,
irrespective of the initial rate of rotation (Burrows 1987; TD93).
Therefore, a coronal driven explosion mechanism is relevant for all
PNSs and relies only upon gravity as the ultimate source of energy.

Observations of core-collapse SNe constrain the value of $f$ for a
given choice of coronal temperature.  Isotropic explosion energies are
measured to be anywhere from $\sim$ a few $\times
10^{50}-10^{52}\,{\rm ergs}$ -- only 0.001-0.1 of the
PNS's gravitational binding energy.  If the corona is purely
responsible for the explosion, then only modest values of
$f\geq 0.001-0.1$ are required.  The upper limit of inferred 
core-collapse SNe
energies derives from so-called ``hypernovae,'' which represent only a
tiny fraction of the overall core-collapse SNe population.  A 
coronal-powered hypernova requires $f\sim 0.1$
and $T\sim$ a few$\times 10^{11}{\rm K}$ to persist for $\sim 1$s 
in order to power a highly energetic event with total kinetic energy
$\sim$ a few$\times 10^{52}\,{\rm ergs}$.

\subsection{Accretion-Powered Gamma-Ray Bursts}

Baryon pollution and the production of large explosion efficiencies
are the two main difficulties which plague models of GRBs (Thompson
1994; Rees 1999).  A coronal-powered explosion mechanism directly
addresses both issues.  For a given energy release,
eq. (\ref{Lsplit2}) dictates that hot diffuse neutrino-emitting
coronae greatly increases the overall explosion efficiency for
adequately large values of $f$.  Thus, models of gamma-ray bursts that
were previously excluded on grounds of energetics may possibly be
reconsidered as valid candidates e.g., mergers of compact objects and
accretion-induced collapse.  A coronal-powered mechanism also
decouples the required intense neutrino flux from the bulk of the
matter, alleviating concerns regarding neutrino induced ablation and
baryon pollution.  An additional feature of a coronal-powered
mechanism is that for a fixed neutrino luminosity, the neutrino
radiation force or momentum deposition rate is larger due to the
increase in opacity, which aids in the collimation of the fireball.

\acknowledgements{Conversations with A. MacFadyen are gratefully
acknowledged.  We also thank S.W. Davis and M. Fukugita for
highlighting the importance of $e-\nu_e$ Comptonization. ERR is
sponsored by NASA through a Chandra Postdoctoral Fellowship award
PF3-40028.  AS acknowledges support of a Hubble Fellowship
administered by the STScI.}

{}

\end{document}

%% file: layer.pstex_t
\begin{picture}(0,0)%
\includegraphics{layer.pstex}%
\end{picture}%
\setlength{\unitlength}{3158sp}%
\begingroup\makeatletter\ifx\SetFigFont\undefined%
\gdef\SetFigFont#1#2#3#4#5{%
  \reset@font\fontsize{#1}{#2pt}%
  \fontfamily{#3}\fontseries{#4}\fontshape{#5}%
  \selectfont}%
\fi\endgroup%
\begin{picture}(7033,6269)(1651,-6361)
\put(4536,-4338){\makebox(0,0)[lb]{\smash{{\SetFigFont{9}{10.8}{\familydefault}{\mddefault}{\updefault}{\color[rgb]{0,0,0}$\nu$}%
}}}}
\put(6327,-2442){\makebox(0,0)[lb]{\smash{{\SetFigFont{17}{20.4}{\familydefault}{\mddefault}{\updefault}{\color[rgb]{0,0,0}$\nu$}%
}}}}
\put(5959,-4338){\makebox(0,0)[lb]{\smash{{\SetFigFont{9}{10.8}{\familydefault}{\mddefault}{\updefault}{\color[rgb]{0,0,0}$\nu$}%
}}}}
\put(7275,-4338){\makebox(0,0)[lb]{\smash{{\SetFigFont{9}{10.8}{\familydefault}{\mddefault}{\updefault}{\color[rgb]{0,0,0}$\nu$}%
}}}}
\put(4726,-6361){\makebox(0,0)[lb]{\smash{{\SetFigFont{11}{13.2}{\familydefault}{\mddefault}{\updefault}{\color[rgb]{0,0,0}turbulent eddies}%
}}}}
\put(5776,-2386){\makebox(0,0)[lb]{\smash{{\SetFigFont{17}{20.4}{\familydefault}{\mddefault}{\updefault}{\color[rgb]{0,0,0}$\nu$}%
}}}}
\put(4651,-2611){\makebox(0,0)[lb]{\smash{{\SetFigFont{17}{20.4}{\familydefault}{\mddefault}{\updefault}{\color[rgb]{0,0,0}$\nu$}%
}}}}
\put(4201,-586){\makebox(0,0)[lb]{\smash{{\SetFigFont{11}{13.2}{\familydefault}{\mddefault}{\updefault}{\color[rgb]{0,0,0}ENERGETIC OUTFLOW}%
}}}}
\put(8251,-3736){\makebox(0,0)[lb]{\smash{{\SetFigFont{14}{16.8}{\familydefault}{\mddefault}{\updefault}{\color[rgb]{0,0,0}{\it H}}%
}}}}
\put(7801,-2461){\makebox(0,0)[lb]{\smash{{\SetFigFont{11}{13.2}{\familydefault}{\mddefault}{\updefault}{\color[rgb]{0,0,0}structure}%
}}}}
\put(7801,-2236){\makebox(0,0)[lb]{\smash{{\SetFigFont{11}{13.2}{\familydefault}{\mddefault}{\updefault}{\color[rgb]{0,0,0}magnetic}%
}}}}
\put(1651,-3511){\makebox(0,0)[lb]{\smash{{\SetFigFont{11}{13.2}{\familydefault}{\mddefault}{\updefault}{\color[rgb]{0,0,0}magnetized }%
}}}}
\put(1651,-3736){\makebox(0,0)[lb]{\smash{{\SetFigFont{11}{13.2}{\familydefault}{\mddefault}{\updefault}{\color[rgb]{0,0,0}corona}%
}}}}
\put(1651,-4786){\makebox(0,0)[lb]{\smash{{\SetFigFont{11}{13.2}{\familydefault}{\mddefault}{\updefault}{\color[rgb]{0,0,0}cool, dense,}%
}}}}
\put(1651,-5011){\makebox(0,0)[lb]{\smash{{\SetFigFont{11}{13.2}{\familydefault}{\mddefault}{\updefault}{\color[rgb]{0,0,0}turbulent}%
}}}}
\put(1651,-5236){\makebox(0,0)[lb]{\smash{{\SetFigFont{11}{13.2}{\familydefault}{\mddefault}{\updefault}{\color[rgb]{0,0,0}layer}%
}}}}
\put(3001,-4336){\makebox(0,0)[lb]{\smash{{\SetFigFont{9}{10.8}{\familydefault}{\mddefault}{\updefault}{\color[rgb]{0,0,0}$\nu$}%
}}}}
\put(1651,-3286){\makebox(0,0)[lb]{\smash{{\SetFigFont{11}{13.2}{\familydefault}{\mddefault}{\updefault}{\color[rgb]{0,0,0}hot, diffuse,}%
}}}}
\put(2401,-2311){\makebox(0,0)[lb]{\smash{{\SetFigFont{11}{13.2}{\familydefault}{\mddefault}{\updefault}{\color[rgb]{0,0,0}hotspot}%
}}}}
\put(3976,-2536){\makebox(0,0)[lb]{\smash{{\SetFigFont{17}{20.4}{\familydefault}{\mddefault}{\updefault}{\color[rgb]{0,0,0}$\nu$}%
}}}}
\end{picture}%